\documentclass[twocolumn,pra,showpacs]{revtex4}
\usepackage{graphicx}
\usepackage[dvips]{color}

\begin{document}

\title{Four-Wave Mixing In BEC Systems With Multiple Spin States}

\author{J. P. Burke Jr.}
\altaffiliation [Present address: ]
{Computer Sciences Corporation, 4090 S. Memorial Pkwy MS-918,
Huntsville, AL 35815.}
\affiliation{100 Bureau Drive, Stop 8423, National
Institute of Standards and Technology, Gaithersburg, MD 20899-8423}
\author{P. S. Julienne}
\affiliation{100 Bureau Drive, Stop 8423, National
Institute of Standards and Technology, Gaithersburg, MD 20899-8423}
\author{C. J. Williams}
\affiliation{100 Bureau Drive, Stop 8423, National
Institute of Standards and Technology, Gaithersburg, MD 20899-8423}
\author{Y. B. Band}
\affiliation{Department of Chemistry, Ben-Gurion University of the Negev,
Beer-Sheva, Israel 84105}
\author{M. Trippenbach}
\affiliation{Institute of Experimental Physics, Optics Division, Warsaw
University, ul.~Ho\.{z}a 69, Warsaw 00-681, Poland}

\date{\today}

\begin{abstract}
We calculate the four-wave mixing (FWM) in a Bose-Einstein condensate
system having multiple spin wave packets that are initially overlapping
in physical space, but have nonvanishing relative momentum that cause
them to recede from one another.  Three receding condensate atom wave
packets can result in production of a fourth wave packet by the process
of FWM due to atom-atom interactions.  We consider cases where the four
final wave packets are composed of 1, 2, 3 and 4 different internal
spin components.  FWM with 1- or 2-spin state wave packets is much
stronger than 3- or 4-spin state FWM, wherein two of the coherent
moving BEC wave packets form a polarization-grating that rotates the
spin projection of the third wave into that of fourth diffracted wave
(as opposed to the 1- or 2-spin state case where a regular
density-grating is responsible for the diffraction).  Calculations of
FWM for $^{87}$Rb and $^{23}$Na condensate systems are presented.
\end{abstract}

\pacs{PACS Numbers: 03.75.Mn, 03.75.Nt, 03.75.Kk} \maketitle

\section{Introduction}

The realization of Bose-Einstein condensation (BEC) of dilute alkali
gases has created considerable interest in the field of nonlinear atom
optics.  In a BEC, pair-wise atomic collisions whose strength is
characterized by the two-body $s$-wave scattering length, $a$, gives
rise to a nonlinear interaction that can produce four-wave mixing
(FWM) phenomena.  The FWM process allows one to study a variety of
phenomena ranging from Bose stimulation \cite{Deng,Tripp98,Tripp00},
elastic scattering loss to empty modes \cite{elcol}, entanglement and
correlations \cite{Vogels}, and squeezing.  The first experimental
demonstration of FWM of matter-waves was reported by Deng, {\em et
al.} \cite{Deng} and involved four BEC wave packets in identical
internal spin states.  The theory of single spin matter-wave FWM was
developed in Refs.~\cite{Tripp98,Tripp00,Meystre99,book}.  But FWM is
also possible for different internal spin states \cite{book}, where
spin exchange collisions may be involved.

This paper makes predictions about the strength of the signal in
multiple spin FWM experiments using a mean field theory with arbitrary
internal spin BEC matter-wave packets, including both elastic and
inelastic loss processes.  FWM can occur in BEC systems containing 1,
2, 3 or 4 different internal spin components.  With 1- or 2-spin
states, the process of FWM is analogous to Bragg diffraction of
matter-waves off a density-grating formed by two of the moving BEC
wave packets, from which the third wave packet can scatter to produce
a fourth wave.  We show that the 3- or 4-spin state case is
phenomenologically different from the 1- or 2-spin state case. In
particular, the nonlinear coupling strength of the former depends on
differences of scattering lengths, greatly reducing the population of
the fourth wave generated by FWM. In this latter case the grating is
no longer a density-grating but a spin-density grating, or
polarization-grating, and the diffraction process rotates the spin
projection of the third wave packet into that of a fourth diffracted
wave packet.

This paper is organized as follows.  Sec.~II provides a general
description of the process of FWM of multispin Bose Einstein
Condensates including the Bragg output coupling technique for
generating high momentum wave packets.  We review general multi-spin
wave packet coupled equations of motion within the Slowly Varying
Envelope Approximation (SVEA) and formulate phase matching conditions
for multi-component spin systems.  Sec.~III describes calculations for
specific cases of 2-, 3- and 4-component cases.  Finally, Sec.~IV
gives a short summary and conclusion.

\section{Theory}

\subsection{Creating a Moving BEC Wave Packet: Initial Conditions}

\begin{figure}[!htb]
\centering
\includegraphics[angle=270,scale=0.4]{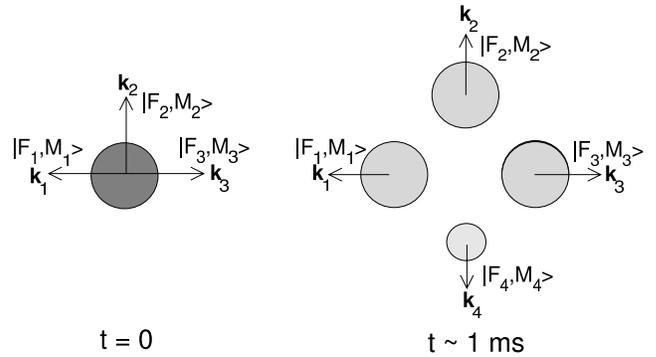}
\caption{Bragg scattering at time $t$=0 creates two moving daughter
wave packets from the original parent condensate.  In the
center-of-mass frame in position space, shown in figure, all three
wave packets are moving with momenta of magnitude $\hbar |k|$ related
to the photon recoil energy.  The nonlinear interaction of the three
initially created wave packets generates a fourth wave packet moving
in the direction satisfying phase-matching criteria.  At $t = 1$ ms
the four wave packets with spin states $|F_j,M_j\rangle$, $j=1,2,3,4$,
have moved away from one another.}
\label{fig1}
\end{figure}

Fig.~\ref{fig1} schematically shows the process of creating moving
BEC wave packets via Bragg scattering wherein two sets of Bragg laser
pulses create two moving daughter wave packets from an original
condensate wave packet.  Two laser pulses of central frequency
$\omega$ and $\omega + \delta$ and wave vector ${\bf k}_{\omega}$ and
${\bf k}_{\omega+\delta}$ incident on a gas of Bose-condensed atoms
impart a well defined momentum ``kick'' to the atoms.  We assume that
the frequency $\omega$ is close to resonance with an atomic
transition, and the detuning frequency $\delta$ is very small compared
with $\omega$.  The frequency $\omega$ is chosen to be close to an
allowed atomic transition (detuned by GHz), and $\delta$ for the pulse
of central frequency $\omega+\delta$ is chosen to be in the kHz
range.
The angle between the propagation directions of the light pulses is
$\theta$ ($\theta = 180^o$ corresponds to the counter-propagating
pulses).  We consider a parent condensate with atoms in a single
internal spin state $\vert F_i,M_i \rangle$.  A set of optical
light pulses incident on the parent condensate with average momentum
$\langle{\bf P}_0\rangle = 0$ can Bragg scatter atoms via Raman
scattering from the initial wave packet, thereby creating a new
daughter wave packet with momentum $\langle{\bf P}\rangle = \hbar({\bf
k}_{\omega} - {\bf k}_{\omega+\delta})$.  This process is associated
with absorption of one photon from the first pulse and stimulated
emission of one photon into the second light pulse.  This is a
resonant transition and occurs only if conservation of energy and
momentum are satisfied simultaneously.  If the resulting velocity
acquired by the atom is large compared to the speed of sound in the
condensate, the dispersion relation is quadratic, and conservation of
energy gives $\hbar \delta = P^{2}/2m + E_{F_f,M_f} - E_{F_i,M_i}$,
where $m$ is the atomic mass, and the subscripts $f$ and $i$ indicate
the final and initial state respectively.  Conservation of momentum
yields the relation $P = 2\hbar k \sin (\theta/2)$, where
$k=2\omega/c$; the atomic recoil momentum equals the difference
between the central momenta of two pulses.  It is often convenient to
specify the velocity of the Bragg scattered atoms in units of the
recoil velocity $v_R \equiv \hbar k/m$ (which equals 2.9 cm/s for
$^{23}$Na).  The intensity and duration of the optical pulses
determine the number of atoms that receive the recoil ``kick'' by
undergoing the Raman process.  The polarization and detuning of the
lasers determine the final internal spin state $\vert F_f,M_f
\rangle$.  Two different sets of Bragg pulses are used to create two
separate daughter wave packets.  These, and what remains of the
initial BEC, are the three nascent wave packets that participate in
the FWM process.

The duration of the Bragg pulses is taken to be short ($\mu$s
time-scale) compared to the mean-field evolution of the BEC (typically
hundreds of $\mu$s).  As a result, it is easy to insure conservation
of energy and momentum in the Raman process that produces Bragg
scattering.  The whole momentum distribution of the initial BEC can
therefore participate in the Bragg scattering process if the Bragg
pulses are sufficiently short.  We therefore use a ``copy''
approximation for the initial conditions of each slowly-varying
envelope created by Bragg scattering.  The three nascent wave packets
can be represented as $\Phi_{r}({\bf x},0) = \sqrt{\frac{N_{r}(0)}{N}}
\Psi_0({\bf x},t=0)$, for $r=1,2,3$ at time $t=0$, $\Phi_{4}({\bf
x},0) = 0$, where $\Psi_0({\bf x},0)$ is the spatial component of the
initial wave function, $N$ is a total number of atoms and $N_{r}(0)$
is the initial number of atoms in the $r$-th component.  Before the
Bragg
pulses are applied, all the atoms are in the same internal spin
state.
If the Bragg pulse sequence triggering the FWM process is associated
with a change in the internal spin state of the atoms, the daughter
wave packet has not only a different momentum than the parent, but
also a different internal spin component.  It is essential that the
Raman detuning from the excited hyperfine state is smaller than the
hyperfine splitting in order for Bragg pulses to change the internal
spin state.

\subsection{SVEA Equations}

Our zero temperature theoretical model for FWM involves condensate
dynamics described by the Gross-Pitaevskii equation (GPE)
\cite{Dalfovo}.  We start our analysis of different multi-spin
component FWM by deriving a set of coupled GPEs for all the
wave packets participating in the process.  The Hamiltonian of the
system in the second quantization can be written as:
\begin{displaymath}
\hat{H} = \sum_\alpha \int \Psi_\alpha^{\dagger}({\bf x},t) \left[
T_{{\bf x}}+
V({\bf x},t))\right] \Psi_\alpha({\bf x},t)d {\bf x}+
\end{displaymath}
\begin{displaymath}
+ \frac{1}{2} \sum_{\alpha,\beta,\gamma,\delta} \int \int
\Psi_\alpha^{\dagger}({\bf x},t) \Psi_\beta^{\dagger}({\bf x'},t)
U_{\alpha\beta,\gamma\delta}({\bf x-x'})
\end{displaymath}
\begin{equation}
\times \Psi_\gamma({\bf x'},t) \Psi_\delta({\bf x},t)d {\bf x}d {\bf
x'}
\label{hamilton}
\end{equation}
where subscripts $\alpha,\beta,\gamma,\delta$ denote different spin
components.  $T_{{\bf x}} = \frac{-\hbar^2}{2m} \nabla_{\bf x}^2$ is
the kinetic energy operator, $V({\bf x},t)$ is an external potential
imposed on the atoms and $U({\bf x-x'})$ is the interaction between
particles, which for the dilute bose gas, is conventionally taken to
be of contact form,
\begin{equation}
U_{\alpha\beta,\gamma\delta}({\bf x-x'})=\frac{4\pi \hbar^2
a_{\alpha\beta,\gamma\delta}}{m}\delta({\bf x-x'}) \,.
\label{CouplingU}
\end{equation}
Here $a_{\alpha\beta,\gamma\delta}$ is the $s$-wave scattering length for
the multispin collision process (see below) and $m$ is the atomic
mass.  The operators $\Psi_\alpha({\bf x},t)$ satisfy equal time
bosonic commutation relations
\begin{equation}
[\Psi_\alpha({\bf x},t),\Psi_\beta({\bf x'},t)] =
[\Psi^\dagger_\alpha({\bf x},t), \Psi^\dagger_\beta({\bf x'},t)] =
0
\end{equation}
\begin{equation}
[\Psi_\alpha({\bf x},t),\Psi^\dagger_\beta({\bf x'},t)]=
\delta({\bf x-x'})\delta_{\alpha\beta}
\end{equation}
The Heisenberg equations of motion for all $\Psi_\alpha({\bf x},t)$
fields can be obtained by taking the commutator with the Hamiltonian
(\ref{hamilton}).  We consider the case when the total wave function
consists of four wave packets moving with different central momenta
${\bf P}_r= \hbar {\bf k}_r, r=1,\ldots,4$.  Within a mean field
approximation we decompose the total wave function into separate
wave packets, centered about momenta ${\bf P}_r$
\begin{equation}
\Psi({\bf x},t) = \sum_{r = 1}^{4} \Phi_r({\bf x},t) \exp(i{\bf k}_r
\cdot {\bf x} - i\omega_r t)|\alpha_{r}\rangle \,.  \label{SVEApsi}
\end{equation}
where the $r$th wave packet with internal spin $\alpha_r$ and mean
kinetic energy $ \hbar\omega_r = \hbar^2 k_r^2/2m$: $\Phi_r({\bf
x},t)$ is the slowly varying envelope (SVE) of packet $r$, $\hbar \bf
k_{r}$ is the central momentum of packet $r$ and
$|\alpha_r\rangle=|F_{r},M_{r}\rangle$ labels the internal atomic spin
state of the wave packet $r$.

Upon substituting the SVE expansion for $\Psi$ in Eq.~(\ref{SVEApsi})
back into the GPE, collecting terms, multiplying by the complex
conjugate of the appropriate phase factors, and neglecting all terms
that are not phase matched, thereby making the slowly varying envelope
approximation (SVEA) \cite{Tripp00}, we obtain a set of coupled
equations for the slowly varying envelopes $\Phi_r({\bf r},t)$.  The
SVEA was necessary to restrict the momentum components only to those
around the central momentum of each of the wave packets in the
numerical calculation; otherwise, the momentum range would have been
too large to treat numerically.  Solving for the SVE allows efficient
numerical simulations and helps in painting a clear picture of the
process since it separates out explicitly the fast oscillating phase
factors representing central momentum $\hbar {\bf k}_r$.  The SVEs
$\Phi_r({\bf x},t)$ vary in time and space on much longer scales than
the phases.  The multicomponent SVEA equations can be written as
\begin{eqnarray}
\left( \frac{\partial}{\partial t} + (-{\bf v}_{r}) \cdot {\bf
\nabla} +
\frac{i}{\hbar}(-\frac{\hbar^{2}} {2m}\nabla^{2} + V_r({\bf x},t) )
\right) \Phi_r & = & \nonumber
\end{eqnarray}
\begin{eqnarray}
-\frac{i}{\hbar}N U_{rr,rr} |\Phi_{r}|^2 \Phi_{r} & & \,\, {\rm
[self}\,\, {\rm phase]}\,\, \nonumber \\
& & \nonumber \\
-\frac{i}{\hbar}N(1+\delta_{\alpha\beta}) \displaystyle{\sum_{s\,(s\ne
r)}}U_{rs,rs} |\Phi_{s}|^2 \Phi_{r} & & \,\,\, {\rm [cross}\,\, {\rm
phase]}\,\, \nonumber \\ & & \nonumber \\ -\frac{i}{\hbar}N
f_{\alpha\beta,\delta\gamma}
\displaystyle{\sum_{s,t,q\,(s \neq t \neq q \neq r)}} U_{qr,st}
\Phi^{*}_{q}\Phi_{s}\Phi_{t} & & \,\,\, {\rm [4WM}\,\,{\rm
source]}\,\, \nonumber \\
\label{SVEA}
\end{eqnarray}
where $\alpha,\beta,\delta,\gamma$ are the respective spin components
of wave packets $q,r,s,t$, and $f_{\alpha\beta,\delta\gamma}$ = 2 if
all spin components are the same~\cite{papererr}(see also
Sec.~\ref{12S} below), = 1 if there are two or four distinct spin
components, and = $\sqrt{2}$ if there are three distinct spin
components.  We assume here that the scattering lengths that enter
into the coupling constants in Eq.~(\ref{CouplingU}) are obtained from
symmetrized scattering matrix elements, as described by Stoof {\it et
al.}~\cite{Stoof88}.

The interaction of the atoms with an external trapping field, which
could be magnetic or optical, results in a harmonic oscillator
potential $V_r({\bf x},t)$ for each spin component.  We allow for a
time dependence of the potential in order to account for the turning
off of the trapping field in the experiments we model.

We only consider the case of zero or extremely weak magnetic field,
such
that the magnetic Zeeman shift is very small compared to the recoil
energy $\hbar k^2/2m$.  In this case, assuming that the kinetic energy
associated with the Bragg `kicks' are much larger than the chemical
potential of the condensate, the momentum and energy conservation
conditions have the form
\begin{eqnarray}
{\bf k}_r + {\bf k}_{q} - {\bf k}_s - {\bf k}_{t} &=& {\bf 0},
\,\,\,\,\,\,\,\,\,\, \label{mom cons} \\
k^2_r + k^2_{q} - k^2_s - k^2_{t} &=& 0 \mathrm{\ \ (in \ CM \ frame)}
\, .
\label{energy_cons}
\end{eqnarray}
Each of the indices $r,q,s,t$ may take values between 1 and 4. If
there were a non-negligible Zeeman splitting in the energies of
different spin-components, then the energy conservation condition
would need to be modified to take into account the energy released in
the inelastic scattering process.  In general, one would expect the
prospects for FWM to be greatly diminshed if the packets have large
relative velocities.

Given our assumption of zero magnetic field, Eqs.~(\ref{mom cons}) and
(\ref{energy_cons}) are automatically satisfied in two cases: (a)
$r=q=s=t$ (all indices are equal), or (b) $r=s \ne q=t$ (two pairs of
equal indices).  The corresponding terms describe what is called in
nonlinear optics self and cross phase modulation terms respectively.
The self and cross phase modulation terms do not involve particle
exchange between different momentum components.  They modify both
amplitude and phase of the wave packet through the mean-field
interaction.

Particle exchange between different momentum wave packets
occurs only when all four indices in the last term on the RHS of
Eq.~(\ref{SVEA}) are different.  In this case, satisfying the
conservation of momentum and energy is not automatic.  4WM can be
viewed as a process in which one particle is annihilated in each wave
packet belonging to an initially populated pair of wave packets and
simultaneously one particle is created in each of two wave packets of
another pair, one of which is initially populated and the other is
initially unpopulated.  Hence, 4WM removes one atom from each of the
``pump'' wave packets (without loosing generality we may call them 1
and 3), and places one atom in the ``probe'' wave packet (we will call
it wave packet 2) and one atom in the 4WM output (wave packet 4).  The
bosonic stimulation of scattering mimics the stimulated emission of
photons in an optical medium.  The phase matching conditions are
particularly simple in the center-of-mass frame, defined by the
conditions $k_1=k_3$, $k_2=k_4$ and $k_2=k_1$.  This picture is a
consequence of the nature of the nonlinear terms in the four SVEA
equations.

For the case of multi-component FWM, there is an additional degree of
freedom that must be included.  In this case there is a coupling
between different internal spin components, introducing new selection
rules namely:
\begin{equation}
F_1=F_2=F_3=F_4 \,\,{\rm or}\,\,F_1=F_2 \neq F_3=F_4 \,,
\end{equation}
and additionally
\begin{equation} \label{msatis}
M_1+M_3=M_2+M_4 \,.
\end{equation}

The initial wave function is obtained from the numerical solution of
the time-dependent GPE using an imaginary time propagation in the
presence of the magnetic potential \cite{Tripp00}.  In the FWM
experiments we model, after turning off the trapping potential, the
BEC wavepacket is allowed to freely evlolve for a time $t_E$ before
the sets of Bragg pulses are applied to produce three initial wave
packets with three different momenta.  To model this, we propagate the
initial wave packet (in real time) for a period of time $t_E$ to
provide the initial condition in Eq.~(\ref{SVEA}).  This free
evolution causes a spatially varying phase to develop across the
condensate as it expands in the absence of the trapping potential due
to the mean-field interaction.  Given the initial condition, the SVEA
equations can be used to propagate the envelope function of each wave
packet, using the same numerical method used to propagate the ordinary
time-dependent GPE.

\subsection{Nonlinear Coupling Constants and Loss Terms}

The goal of this paper is to estimate the number of atoms in the
created FWM wave packet in various realizations of the multi-spin FWM
experiments.  To accomplish this goal we must first determine the
nonlinear coupling constants $U_{qr,st}$ for $^{23}$Na and $^{87}$Rb
in the $F=1$ and $F=2$ internal spin states that appear in the GPE. We
performed the calculations of the strength of the nonlinear coupling
constant $U_{rq,st}(k_{ts})$, which is determined by the following
two-body scattering process
\begin{equation}
|F_{s},M_{s}\rangle + |F_{t},M_{t}\rangle \rightarrow
|F_{q},M_{q}\rangle + |F_{r},M_{r}\rangle \,,
\end{equation}
at relative momentum $k_{st} = |{\bf k}_{s} - {\bf k}_{t}|$.  The
nonlinear coupling constant is defined by $U_{qr,st}(k_{st})/
\hbar=\frac{4\pi\hbar}{m}\left(-T_{qr,st}/k\right)$, where $S=1+2iT$
is the
unitary
scattering matrix.  For the {\em elastic collision} case, and for
small values of relative momentum $k_{st}$, the coupling constant is
\begin{equation}\label{eloss2}
\frac{-T_{st,st}}{k_{st}} \approx a_{st} \equiv A_{st}-iB_{st}
\, ,
\end{equation}
where $a_{st}$ is the complex $s$-wave scatttering
length~\cite{Julienne02}.  The total elastic cross section is given by
$\sigma_{{\rm elastic}} = (1+\delta_{st}) 4 \pi (A_{st}^2+B_{st}^2)$,
and the total rate constant for inelastic collisions is $(K_2^{{\rm
total}})_{st} = \sum_{qr} (K_2)_{st \rightarrow qr}={4h \over
m}B_{st}$.  The elastic and inelastic collisional losses from the
moving wave packets can be calculated using $\sigma_{{\rm elastic}}$
and $K_2$, as described in Ref.~\cite{elcol}.  We calculate the
various $A_{st}$ and $B_{st}$ values for $^{23}$Na and $^{87}$Rb
collisions using standard coupled channels models of threshold
scattering.

Insight into the nature of the four-wave mixing source terms can be
obtained by using a scattering representation that is useful at low
magnetic field, where the total angular momentum
${\bf F}_t={\bf F}_1+{\bf F}_3+{\bf l}= {\bf f}+{\bf l}$ of the
colliding pair of atoms is conserved in a collision.  The T-matrix
elements we need are for the case when two atoms collide in the
internal spin levels ($F_1,M_1$), ($F_3,M_3$) and with relative
angular momentum $\ell,m$ and end up in the levels($F_2,M_2$),
($F_4,M_4$) with relative angular momentum $\ell',m'$.  We assume
collision energy is low enough that only $s$-waves contribute, so that
$\ell=\ell'=0$, $m=m'=0$.

To obtain the relevant quantum numbers, first vector couple the
$|F_1M_1\rangle$ and $|F_3M_3\rangle$ levels to obtain the resultant
angular momentum $|F_1F_3f,M_1+M_3\rangle$ states, then vector couple
the relative angular momentum $|\ell,m\rangle$ to get the total
angular momentum states $|F_1F_3f,\ell,F_t, M_t=M_1+M_3+m\rangle$.
The desired symmetrized $T$-matrix elements~\cite{Stoof88} are given
in terms of the indices $\{F_1F_3f \ell F_t\}$ and are independent of
projection quantum numbers $M_1,M_3, m, M_t$.

The only significant collisions we need to consider are spin-exchange
collisions.  These are possible only if the following selection rules
are obeyed: $\ell=\ell'$, $m=m'$, $f=f'$, and $M_1+M_3=M_2+M_4$; for
$s$-waves $\ell=0$ and $f'=f=F_t$.  The four-wave mixing source term
is proportional to $a(st\to qr)=-T_{qr,st}/k_{st}$.  Using the
transformation of $T$-matrix elements to the total angular momentum
basis~\cite{Mies73,Baylis87,Suominen98}, assuming $\ell=\ell'=0$ we
obtain
\begin{eqnarray}
&& a(F_1M_{1},F_3M_3 \to F_2M_2,F_4M_4) =\nonumber\\
&& \left(\frac{1+\delta_{F_1F_3}}{1+\delta_{F_1F_3}\delta_{M_1M_3}}
\right)^{\frac{1}{2}} \left(\frac{1+\delta_{F_2F_4}}{1+\delta_{F_2F_4}
\delta_{M_2M_4}} \right)^{\frac{1}{2}} \nonumber\\
&& \times \sum _{f} (F_2F_4f|M_2M_4,M_2+M_4) \nonumber\\ && \times
(F_1 F_3 f|M_1M_3,M_1+M_3) a^{(f)}(F_1F_3 \to F_2F_4) ,\nonumber\\
\label{rf.4a}
\end{eqnarray}
where $(\ |\ )$ are Clebsch-Gordan coefficients.  Clearly four-wave
mixing is possible only if the collisions are energy degenerate:
$F_2=F_1$ and $F_4=F_3$.

For 1- and 2-component four wave mixing, where $M_1=M_3=M_2=M_4$ or
$M_2=M_1$ and $M_4=M_3$ respectively, the product of Clebsh-Gordon
coefficients is positive definite, and the sum will be of the same
magnitude as the individual elements.  On the other hand, for three-
or four-component four wave mixing, there will be terms with the
product of the Clebsh-Gordon coefficients being positive and terms
with the product being negative.  In this case, the scattering length
that controls the FWM source term in Eq.~(\ref{SVEA}) will involve
{\em differences} of scattering lengths of comparable magnitudes, and
will tend to be much smaller than for the one-or two-component case.
For example,
\begin{eqnarray}
&&a(1-1,11 \to 10,10) = \nonumber \\
&&\frac{\sqrt{2}}{3}\left(a^{(2)}(11\to11) - a^{(0)}(11\to11) \right)
\, , \\
&&a(1-1,2-1 \to 10,2-2) = \nonumber \\
&&\frac{\sqrt{2}}{3}\left(a^{(3)}(12\to12) - a^{(2)}(12\to12) \right)
\, .
\label{difference}
\end{eqnarray}
Inelastic energy-releasing exchange collisions are only possible for
the cases$\{F_1F_3f\}=$ $\{122\}$, $\{220\}$ and $\{222\}$ for
$^{23}$Na and $^{87}$Rb.  For $^{87}$Rb, these loss collisions are
anomously small \cite{julienne97,burke97,kokkelmans97} and we can set
$B^{(f)}(F_1F_3)=0$.

Table \ref{table1} gives our calculated scattering lengths, based on
coupled channels models of threshold collisions, that should be used
in the $U_{qr,st}$ coupling constant that gives the source term for
FWM in Eq.~(\ref{SVEA}), i.e., $U_{qr,st} = 4\pi a_{qr,st}
\hbar^{2}/m$.  We include several cases for $^{23}$Na and $^{87}$Rb
involving one, two, three, and four spins.  The three and four spin
cases have significantly larger coupling constants for $^{23}$Na than
for $^{87}$Rb, because scattering length differences in
(\ref{difference}) tend to be much larger for the former case.  The imaginary
part of the complex scattering length for the source term is small
compared to the real part, even for $^{23}$Na collisions, and is not shown.

\section{Result of Calculations}

\subsection{One- And Two-Spin Component FWM} \label{12S}

In the numerical simulations we model the condensate comprised of
magnetically trapped atoms without a discernible non-condensed
fraction.  We assume trap frequencies in the $x$, $y$ and $z$
directions of 84, 59 and 42 Hz, as in Ref.\cite{Deng}.  After the
magnetic trap is switched off, the condensate expands freely during a
delay time $t_E$.  Then the sequence of two sets of Bragg pulses
creates two moving wave packets 2 and 3 in the lab frame, in addition
to the initial stationary wave packet 1.  One can change the momentum
of the wave packets by changing the angle between the laser beams used
for the Bragg out-coupling or by changing the laser frequencies.  Our
simulations neglect the detailed dynamics during the application of
the
Bragg pulses; instead, we assume that after the Bragg pulses are
applied, each wave packet is a copy of the parent condensate wave
function.  We shall consider the case where the initial three wave
packets contain an equal number of atoms (except for the case
corresponding to the original NIST experiments where the ratio is
3:7:3).

In the center-of-mass reference frame, all the wave packets (including
the new FWM wave packet) move with the same velocity.  In the one and
two-spin component cases the thresholds of the incoming and outgoing
collision pairs coincide, i.e., $F_{q} + F_{r} = F_{s} + F_{t}$.
Moreover, $M_{F}$ is a good quantum number in the two-body collision
requiring $M_{q}+M_{r} = M_{s} + M_{t}$ for all cases.  The three
initial wave packets can each be in an arbitrary internal spin state
of the alkali atom (for $^{23}$Na or $^{87}$Rb this is either $F=1$ of
$F=2$).  The FWM process preferentially populates the fourth wave
packet in a spin state that satisfies the energy and angular momentum
projection constraints.

Of all the possible combinations of spin states we can start with, we
find two distinct classes of combinations: (1) at least two of the
three initial wave packets are in the same spin state, (2) all three
wave packets are in different spin states.  In both cases, the overlap
of the coherent moving BEC packets, 1 and 2, form a grating (either a
density-grating or a spin-density grating) that 3 diffracts off,
producing a new wave packet 4.  In the single-spin case the number of
atoms in the fourth wave packet is four times larger than in the
2-spin case, $(N_{4})_{\rm 1-spin} \sim 4(N_{4})_{\rm 2-spin}$ even if
the scattering lengths and initial number of atoms for these cases
were exactly identical.  The grating picture that explains the factor
of four is as follows.  For the 1-spin state case, wave packets 1 and
2 form a grating and wave packet 3 can scatter off the grating to
produce atoms that are in wave packet 4 (see Fig.~\ref{fig1}), {\em
and} wave packets 2 and 3 form a grating and wave packet 1 can scatter
off the grating to produce atoms that are in wavepacket 4.  These two
amplitudes {\em add up coherently} so the probability is four times
the probability that would be obtained were there only one amplitude
for producing the 4th wave.  For the two-spin state case, if packet 1
(or 3) is the different spin state, it will scatter off the grating
formed by 2 and 3 (1 and 2) and form the same spin state in a wave
packet with momentum ${\bf k}_4$.  The gratings are dynamical ones
that change in time as atoms are removed from wave packets 1 and 3 and
placed in wave packets 2 and 4.

\begin{figure}[!htb]
\centering
\includegraphics[angle=270,scale=0.4]{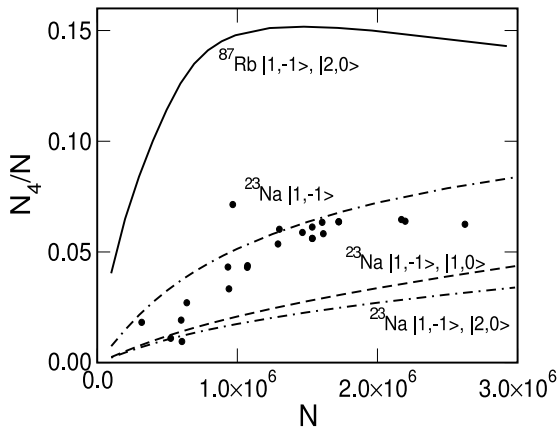}
\caption{Fractional FWM output versus total number of atoms $N$, $f_4 =
N_4/N$, in cases when one- or two-spin components are present.  The
simulations were carried out allowing the condensate to expand for 600
$\mu$s before applying the Bragg pulses, as in the NIST experiment
\cite{Deng}.  The solid circles are the experimental results obtained
for $^{23}$Na $|1,-1\rangle$.  In the two-spin component simulations
the ratio of atoms in the three initially populated wave packets are
1:1:1, whereas the one-spin component simulation used the experimental
ratios of 7:3:7.}
\label{fig2}
\end{figure}

The $s$-wave scattering lengths were used to form the nonlinear
coupling constants, $U_{qr,st} = 4\pi a_{qr,st}\hbar^{2}/m$, and in
turn, these were used in our numerical simulations of FWM.
Fig.~\ref{fig2} shows the fraction of atoms in the newly generated
wave packet, $f_4 = N_4/N$, versus the total number of atoms in the
parent condensate $N$ for a number of different FWM processes.  We
assumed a free expansion time (i.e., the delay time) of $t_E=600$
$\mu$s, as in the NIST experiment~\cite{Deng}.  The strongest FWM
conversion is obtained for $|1,-1\rangle_{1} + |2,0\rangle_{3}
\longrightarrow |1,-1\rangle_{2} + |2,0\rangle_{4}$ collisions of
$^{87}$Rb atoms, since this case has the largest $s$-wave scattering
length in the source term.  The saturation (and even decrease in the
$^{87}$Rb case) of the FWM fraction $f_4 \equiv N_4/N$ with increasing
$N$ is clear from Fig.~\ref{fig2}.  The origin of this saturation is discussed in
Ref.~\cite{Tripp00} and is due to the physical separation of the interacting
wave packets,  back conversion from the newly formed wave packet 4, and
elastic and inelastic scattering loss processes.

For reference, we included in Fig.~\ref{fig2} the original data from
the NIST experiment \cite{Deng}.  Comparison of the experimental
results for $^{23}$Na $|1,-1\rangle$ with the theoretical results show
that at the highest values of $N$, the calculated number of atoms in
the FWM wave packet is higher than in experiment.  It is possible that
stimulated elastic scattering loss may have to be taken into account
at large $N$.  \\
\bigskip

\begin{figure}[!htb]
\centering
\includegraphics[angle=0,scale=0.30]{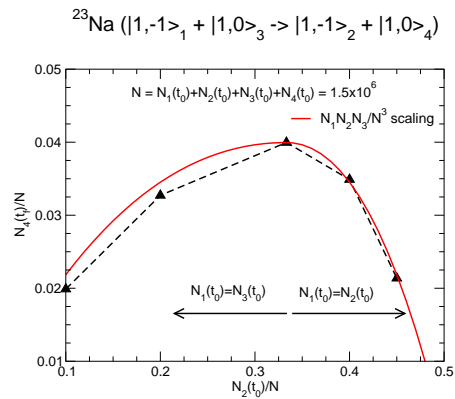}
\caption{Fractional FWM output versus $N_2(t_0)/N$.  The solid curve is
the result of using the simple model in Eq.~(\ref{model}) and the
triangles are the result of calculations.     The trap is the same as used
in Fig.~\ref{fig2}.    The Bragg pulses are applied 300 $\mu$s after the
trapping potential is turned off.}
\label{fig3}
\end{figure}

The scaling of the fraction of atoms output into the FWM wave packet
with $N_2(t_0)/N$ for $N = N_1(t_0) + N_2(t_0) + N_3(t_0) = N =
1.5\times 10^6$ is shown in Fig.~\ref{fig3} for the case of
$|1,-1\rangle_{1} + |1,0\rangle_{3} \longrightarrow |1,-1\rangle_{2} +
|1,0\rangle_{4}$ collisions of $^{23}$Na atoms.  Note that on the left
part of the figure $N_1(t_0) = N_3(t_0) > N_2(t_0)$, while on the
right part of the figure $N_1(t_0) = N_2(t_0) > N_3(t_0)$.  In
Ref.~\cite{Tripp00}, we discussed how the fraction of atoms in the FWM
wave packet would scale with the initial number of atoms in the three
nascent wave packets.  There we showed how a simple argument predicts
that the total FWM output fraction is given by
\begin{equation} \label{model}
    f_4 = \frac{N_4(t_{col})}{N} \approx f_1 f_2 f_3 \left
    (\frac{t_{col}}{t_{NL}} \right)^2 \,,
\end{equation}
where $f_i = N_i(t_0)/N$, $t_{col}$ is the time for the wave packets
to separate, and $t_{NL}$ is the characteristic nonlinear time scale,
$t_{NL}=\hbar/\mu$ where $\mu$ is the chemical potential.  This is an
upper bound on the 4WM output, since the mutual interaction of the
packets due to the self- and cross-phase modulation terms (the self-
and cross-interaction energy terms), and the elastic and inelastic
loss processes are not included in the estimate.  The curve labeled
$N_1 N_2 N_3/N^3$ in Fig.~\ref{fig3} is obtained using
Eq.~(\ref{model}).  It follows the calculated results rather well.
Clearly, for cases where $N < 1.5\times 10^6$, one expects that the
simple argument will provide as good an estimate.

\begin{figure}[!htb]
\centering
\includegraphics[angle=270,scale=0.4]{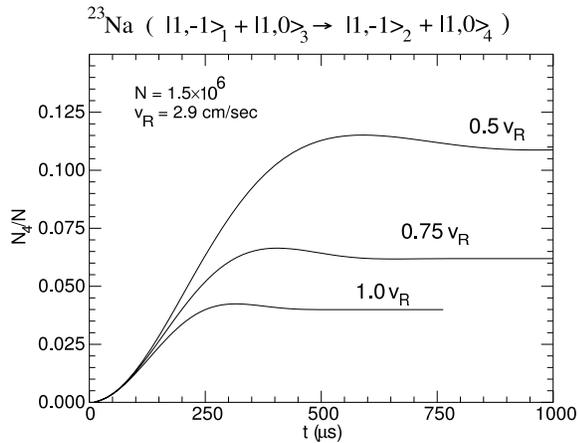}
\caption{$N_{4}(t)/N$ versus $t$ for $N = 1.5\times 10^6$ atoms for
various velocities $\hbar k/m$ of the wave packets.  The trap is the
same as used in Fig.~\ref{fig2}.  The Bragg pulses are applied 300
$\mu$s after the trapping potential is turned off.}
\label{fig4}
\end{figure}

The dependence of the fraction of atoms in the FWM wave packet on the
velocity of the condensate wave packets is shown in Fig.~\ref{fig4}
for the case of $^{23}$Na $|1,-1\rangle_{1} + |1,0\rangle_{3}
\longrightarrow |1,-1\rangle_{2} + |1,0\rangle_{4}$ condensate
collisions where the free-expansion time $t_E$ between trap off and
the Bragg pulses is 300 $\mu$s.  The velocities of the wave packets in
the center-of-mass frame are indicated in Fig.~\ref{fig4} in units of
the recoil velocity $v_R$ (recall that the recoil velocity is defined
as $v_R = \hbar k/m$ which equals 2.9 cm/s for $^{23}$Na).  The
relative velocities of wave packets 1 and 3 or 2 and 4 are twice these
velocities.  As the relative velocity increases, the number of atoms
in the FWM packet decreases, since the duration of the wave packet
overlap essential to FWM decreases.  For example, for the curve labeled
$1.0v_R$ in Fig.~\ref{fig4} it takes about 500 $\mu$s for the centers of
the moving wave packets to separate by the mean Thomas-Fermi
diameter of the initial condensate.  The saturation of $N_4/N$ is
evident.  For very slow relative velocities, $N_4/N$ actually begins
to decrease with time due to back-conversion via FWM, as first
described in Ref.~\cite{Tripp00}.

\begin{figure}[!htb]
\centering
\includegraphics[angle=270,scale=0.4]{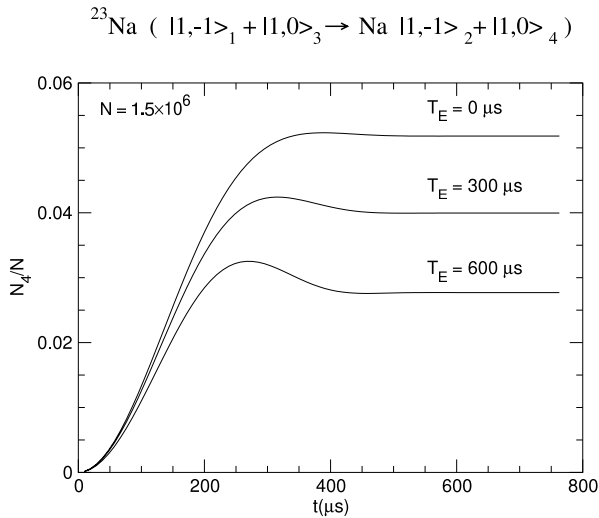}
\caption{$N_{4}(t)/N$ versus $t$ for $N = 1.5\times 10^6$ atoms for
various free-expansion times $t_E$.  The trap frequencies are the same
as used in Fig.~\ref{fig2}. }
\label{fig5}
\end{figure}

The dependence of the fraction of atoms in the FWM wave packet on the
free-expansion time $t_E$ between turning off the harmonic potential
and applying the Bragg pulses is shown in Fig.~\ref{fig5} for the case
of $^{23}$Na $|1,-1\rangle_{1} + |1,0\rangle_{3} \longrightarrow
|1,-1\rangle_{2} + |1,0\rangle_{4}$ condensate collisions when the
wave packets velocity is $v_R = 2.9$ cm/s.  During
the free-expansion of the condesate, a spatially varying phase
develops across the parent condensate, and this phase deleteriously
affects the phase matching required for FWM. This has already been
discussed in the single-component studies that were published
previously \cite{Tripp00}.

\begin{figure}[!htb]
\centering
\includegraphics[scale=0.25]{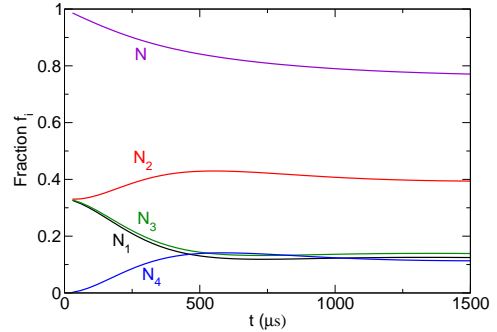}
\caption{$N(t)$, $N_{1}(t)$, $N_{2}(t)$, $N_{3}(t)$, and $N_{4}(t)$
versus $t$ for $^{87}$Rb $|1,-1\rangle _1 + |2,0\rangle _3
\longrightarrow |1,-1\rangle _2 + |2,0\rangle _4$ collisions with $N =
1.0\times 10^6$ atoms.  The trap frequencies are the same as used in
Fig.~\ref{fig2}.  The Bragg pulses are applied 600 $\mu$s after the
trapping potential is turned off. }
\label{fig6}
\end{figure}

Figure \ref{fig6} shows the total number of atoms in all the wave
packets and the number of atoms in each of the wave packets during the
FWM half-collision versus time for the $^{87}$Rb $|1,-1\rangle _1 +
|2,0\rangle _3 \longrightarrow |1,-1\rangle _2 + |2,0\rangle _4$ case.
A substantial loss of the condensate atoms due to elastic and
inelastic scattering collisions occurs, but nevertheless the newly
created FWM wave packet contains 120,000 atoms after full separation.
This number of atoms can be easily detected.  In fact, for all the
cases shown in Fig.~\ref{fig2}, the generated FWM signal is strong
enough to be detected in real experiments when $N \geq 1.0\times 10^6$
atoms.

\subsection{Three- and Four-Spin Component FWM}

FWM is possible with any combination of arbitrary internal spin states,
provided $M_F$ is conserved in the elastic two-body scattering process.
FWM processes cannot be interpreted as Bragg diffraction off
density-gratings in 3- or 4-spin FWM. Instead, a
``polarization-grating'' scatters the atoms, as described above.  For
example, the four-mixing output (packet 4) in the process
\begin{equation}
|1,-1\rangle_{1} + |1,+1\rangle_{3} \rightarrow |1,0\rangle_{2} +
|1,0\rangle_{4}
\end{equation}
can be thought of as a rotation of the spin ``polarization'' of wave
packet 3 due to scattering off the spin density grating formed by the
overlap wave packets 1 and 2.

\begin{figure}[!htb]
\centering
\includegraphics[angle=0,scale=0.4]{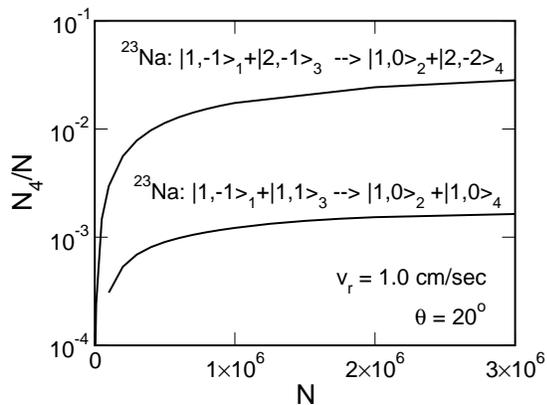}
\caption{Fractional FWM output versus total number atoms $N$ calculated
upon allowing the condensate to expand for 600 $\mu$s before applying
the Bragg pulses for $^{23}$Na $|1,-1\rangle_{1} + |1,+1\rangle_{3}
\rightarrow |1,0\rangle_{2} + |1,0\rangle_{4}$ and for $^{23}$Na
$|1,-1\rangle_{1} + |2,-1\rangle_{3} \rightarrow |1,0\rangle_{2} +
|2,-2\rangle_{4}$. The initial ratio of atoms in the three populated
wave packets are 1:1:1.   The trap frequencies are the same as used in
Fig.~\ref{fig2}.  }
\label{fig7}
\end{figure}

\begin{figure}[!htb]
\centering
\includegraphics[scale=0.3]{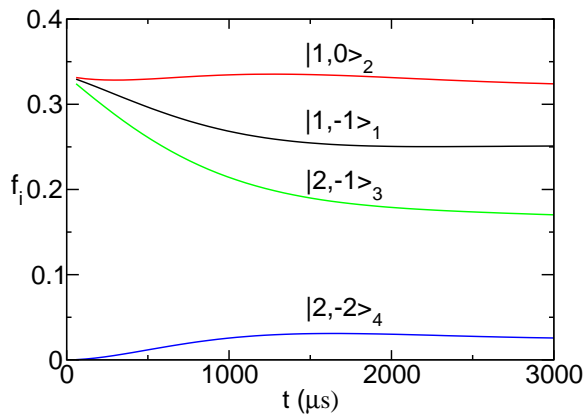}
\caption{$N_{1}(t)$, $N_{2}(t)$, $N_{3}(t)$, and $N_{4}(t)$
versus $t$ for $N = 2\times 10^6$ atoms and $^{23}$Na
$|1,-1\rangle_{1} + |2,-1\rangle_{3} \rightarrow |1,0\rangle_{2} +
|2,-2\rangle_{4}$.   The trap frequencies are the same as used in
Fig.~\ref{fig2}.  The Bragg pulses are applied 600 $\mu$s after the
trapping potential is turned off. }
\label{fig8}
\end{figure}

Fig.~\ref{fig7} shows $N_4/N$ versus $N$ for $^{23}$Na
$|1,-1\rangle_{1} + |1,+1\rangle_{3} \rightarrow |1,0\rangle_{2} +
|1,0\rangle_{4}$ and for $^{23}$Na $|1,-1\rangle_{1} +
|2,-1\rangle_{3} \rightarrow |1,0\rangle_{2} + |1,-2\rangle_{4}$
condensate collisions.  Fig.~\ref{fig8} shows the time-dependent
population fractions for the latter 4-component case for $N =
2.0\times 10^6$ atoms.  The $s$-wave scattering length in the source
term for the former process is $a = 0.14$ nm, whereas for the latter
process, $a = 0.59$ nm.  Taking an angle of 20 degrees between the
Bragg laser pulses ensures a relatively low velocity of $\hbar k/m=$ 1
cm/sec for the moving packets.  This allows the packets to remain
overlapped long enough to generate an observable fourth wave.

\section{Summary and conclusions}

We have developed a general theory for describing four-wave mixing
(FWM) of matter-waves in arbitrary internal spin states within the
context of a mean-field theory using the Gross-Pitaevskii equation
(GPE).  The slowly varying envelope approximation is used to write
separate equations for each of the condensate wave packets.  The
atom-atom interactions of the BEC atoms result in both the mean-field
interaction terms, analogous to the self and cross phase modulation
terms in nonlinear optics, and the FWM source terms in the GPE. The
theory also incorporates elastic and inelastic scattering loss
processes.  These processes take atoms out of the condensate wave
packets and therefore reduce the FWM. FWM with 1- or 2-spin states is
analogous to Bragg diffraction of matter-waves off a density-grating
formed by the moving BEC wave packets.  The 2-spin state FWM is
generally smaller than the 1-spin state case by a factor of about four
due to the coherent addition of two amplitudes for scattering off the
density gratings formed in the 1-spin case, whereas only one
scattering amplitude occurs for the 2-spin state case.  FWM with 3- or
4-spin state are generally much weaker processes; in the 3- or 4-spin
state cases, the coherent moving BEC wave packets form a
polarization-grating (a spin-density grating) that rotates the spin
projection of the diffracted wave.

Calculations of multicomponent FWM for $^{87}$Rb and $^{23}$Na
condensate systems have been presented.  In these calculations the 3-
and 4-spin state FWM output signals are lower by roughly an order of
magnitude than for 1- and 2-spin state FWM cases.  The reduction is
due to the much smaller source term for FWM in the 3- and 4-spin state
cases, since the coupling strength involves differences of scattering
lengths (see Eq.~(\ref{difference})) of comparable magnitudes.

\bigskip

\begin{acknowledgments}
This work was supported in part by grants from the Office of Naval Research, the National
Research Council, the U.S.-Israel Binational Science Foundation (grant No.~2002147), and
the Israel Science Foundation for a Center of Excellence Grant (grant No.~212/01) and KBN
grant 2P03B04325 and Polish Ministry of Scientific Research and Information Technology
under grant PBZMIN- 008/P03/2003 (M.T.).
\end{acknowledgments}

\begin{table}
\caption{Scattering lengths $a_{qr,st}$ in nm for the source nonlinear
coupling constant $U_{13,24}$.}
\begin{tabular}{|l|c|c|} \hline
Spin States & $^{23}$Na &  $^{87}$Rb \\ \hline\hline
 1 & & \\ \hline
$|1,-1\rangle_{1} + |1,-1\rangle_{3} \rightarrow
|1,-1\rangle_{2} + |1,-1\rangle_{4}$ & 2.89 & 5.63 \\ \hline
 2 & & \\ \hline
$|1,-1\rangle_{1} + |1,0\rangle_{3} \rightarrow
|1,-1\rangle_{2} + |1,0\rangle_{4}$ & 2.89 & 5.63 \\ \hline
$|1,-1\rangle_{1} + |2,0\rangle_{3} \rightarrow
|1,-1\rangle_{2} + |2,0\rangle_{4}$ & 2.78 & 5.51  \\ \hline
 3 & & \\ \hline
$|1,-1\rangle_{1} + |1,+1\rangle_{3} \rightarrow |1,0\rangle_{2} +
|1,0\rangle_{4}$ & 0.14 & -0.01 \\ \hline
$|2,+1\rangle_{1} + |2,-1\rangle_{3} \rightarrow |2,0\rangle_{2} +
|2,0\rangle_{4}$ & 0.64 & 0.38 \\ \hline
$|2,+2\rangle_{1} + |2,0\rangle_{3} \rightarrow |2,1\rangle_{2} +
|2,1\rangle_{4}$ & 0.78 & 0.44 \\\hline
  4 & & \\ \hline
$|1,-1\rangle_{1} + |2,-1\rangle_{3} \rightarrow |1,0\rangle_{2} +
|2,-2\rangle_{4}$ & 0.59 &0.09 \\ \hline
$|1,-1\rangle_{1} + |2,0\rangle_{3} \rightarrow |1,0\rangle_{2} +
|2,-1\rangle_{4}$ & 0.36 &0.06 \\ \hline
$|2,+1\rangle_{1} + |2,0\rangle_{3} \rightarrow |2,2\rangle_{2} +
|2,-1\rangle_{4}$ & 0.35 &0.22 \\ \hline
\end{tabular}
\begin{flushleft}
The scattering lengths are given to the nearest 0.01 nm, although
these numbers are not necesaarily accurate to 0.01 nm.  The small
numbers, depending on differences of scattering lengths as in
Eq.~(\ref{difference}), are, in particular, subject to revision as
improved threshold scattering models develop based on the most recent
high quality experimental data.
\end{flushleft}
\label{table1}
\end{table}


\begin{thebibliography}{}
\bibitem{papererr} The factor of two was correctly used in the
calculations reported in Ref.~\cite{Tripp00}, although inadvertently
omitted from the source term on the right hand side of the printed
Eqs. (9) and (12) - (15) in that paper.
\end{thebibliography}
\end{document}